\documentclass[twocolumn,amsmath,amssymb]{revtex4}
\usepackage{graphicx}
\usepackage{color}

\begin{document}

\title{Difficult Sudoku Puzzles Created by Replica Exchange Monte Carlo Method}

\author{Hiroshi Watanabe}

\affiliation{
The Institute for Solid State Physics, The University of Tokyo,
Kashiwanoha 5-1-5, Kashiwa, Chiba 277-8581, JAPAN
}

\date{\today}

\begin{abstract}
An algorithm to create difficult Sudoku puzzles is proposed.
An Ising spin-glass like Hamiltonian describing  difficulty of puzzles is defined,
and difficult puzzles are created by minimizing the energy of the Hamiltonian.
We adopt the replica exchange Monte Carlo method with simultaneous temperature adjustments
to search lower energy states efficiently, and 
we succeed in creating a puzzle which is the world hardest ever created in our definition,
to our best knowledge.
(Added on Mar. 11, the created puzzle can be solved easily by hand. 
Our definition of the difficulty is inappropriate.)
\end{abstract}

%\pacs{02.50.Ng ,05.50.+q, 89.75.Da}

\maketitle

\section{Introduction}

\begin{figure}
\begin{center}
\includegraphics[width=4.1cm]{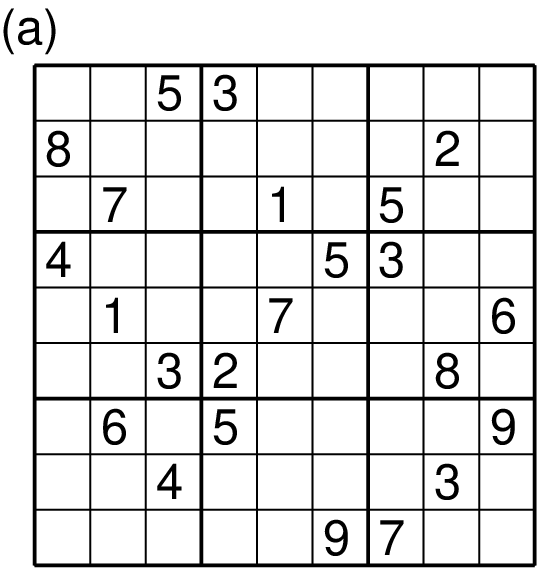}
\includegraphics[width=4.1cm]{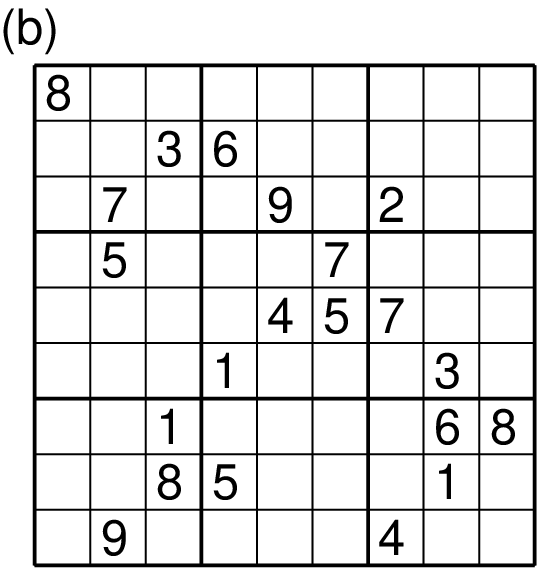}
\end{center}
\caption{Sudoku puzzles created by Dr. Inkara.
(a) The puzzle made in 2010 (Inkara2010), which depth is 5, normal width is 173,
and average width 179 $\pm$ 3.25, respectively.
(b) The puzzle made in 2012  (Inkara2012), which depth is 8, normal width is 3599,
and average width = 2257 $\pm$ 25.7, respectively.
}\label{fig_inkara}
\end{figure}

Sudoku, which is also called Number Place, is a kind of pencil puzzles~\cite{Nikoli}.
Each Sudoku puzzle has $9 \times 9$ grid, which has nine $3 \times 3$ subgrids.
There are some numbers in cells.
The objective of the puzzle is to complete the grid 
by filling numbers in empty cells so that
each row, column, and subgrid contains all of numbers from 1 to 9.
Sudoku puzzles are now extremely popular in the world, and recently
Sudoku have attracted much attention as mathematical and physical points of view.
In 2002, solving Sudoku puzzle is proved to be NP-complete problems~\cite{NPcomplete}.
All possible Sudoku solutions is enumerated by Felgenhauer and Jarvis~\cite{Felgenhauer}.
Solving Sudoku puzzles corresponds to find the grand state
of the antiferromagnetic 9-state Potts model with special interactions.
The similarity between the Sudoku problems and spin-glass systems has been pointed out~\cite{TanakaHukushima}.
Williams and Ackland defined a Sudoku Hamiltonian, and observed thermodynamic phase transitions
by utilizing Monte Carlo (MC) simulations~\cite{Williams2012}.
They also pointed out that the energy landscape of
the Sudoku Hamiltonian is rugged, and the model show similar behavior to spin-glass systems.
From the view point of the computing, it is rather easy to find the solution of the given puzzle.
However, it is not trivial to make difficult puzzles by using computers.
In 2010, Dr.~Arto Inkara created a difficult Sudoku puzzle (Inkara2010)~\cite{Inkara2010}
which is shown in Fig.~\ref{fig_inkara}~(a).
Later, he created a more difficult one in 2012 (Inkara2012)~\cite{Inkara2012} which is shown in Fig.~\ref{fig_inkara}~(b).
To our best knowledge, Inkara2012 is the world hardest Sudoku puzzle ever created.
The purpose of the present manuscript is to create a Sudoku puzzle which is more difficult than Inkara2012,
and consequently is the world hardest, by utilizing a MC method.

This manuscript is organized as follows.
A method to solve Sudoku puzzles and a definition of difficulty are described in Sec.~\ref{sec_difficulty}.
The algorithms for finding difficult puzzles are explained in Sec.~\ref{sec_method}.
The numerical results are given in Sec.~\ref{sec_results} and 
a summary and a discussion of further issues are given in Sec.~\ref{sec_summary}.

\section{Definition of Difficulty} \label{sec_difficulty}

\begin{figure}
\begin{center}
\includegraphics[width=8cm]{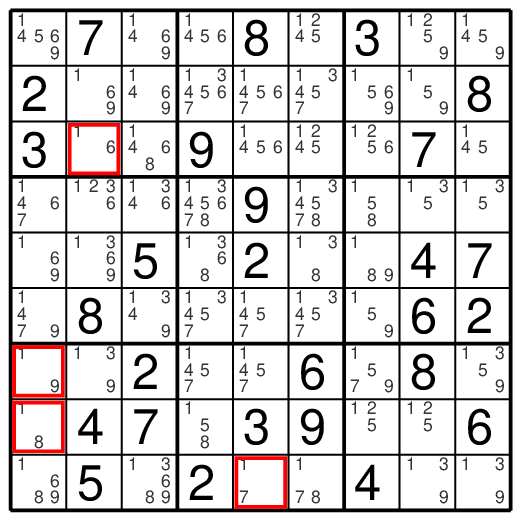}
\end{center}
\caption{
(Color online) Pencil marks in sudoku cells. 
Each empty cell contains small numbers which are possible inside the cell.
The small numbers are called pencil marks.
If an empty cell has only one pencil mark, that mark is the value of the cell.
In this puzzle, the smallest number of pencil marks in a cell is two.
There are four cells which have two candidates.
}\label{fig_pm}
\end{figure}

Difficulty of Sudoku puzzles depends on a method to solve them.
Therefore, we describe the method to solve Sudoku puzzles.
While there are many kinds of techniques, we adopt only two of them,
pencil marks and recursive backtracking (see Fig.~\ref{fig_pm}).

\textbf{Pencil Marks}: Pick up an empty cell.
Check all the numbers in the row, column, and subblock to which the cell belongs.
Then list up all numbers which are still possible in the empty cell.
These numbers are called pencil marks.
If a cell has only one pencil mark, then the mark is the value of the cell.
Repeat the above procedure until all empty cells have two or more pencil marks.

\textbf{Recursive Backtracking}: 
Pick up the cell which has the smallest number of pencil marks.
Then choose one of the pencil marks and assume that it is the value of the cell, and continue to solve the problem recursively.
If the assumed value does not lead to the complete solution, then choose another value of the pencil marks.

The algorithm to solve Sudoku puzzles involves the iteration of pencil marks
and recursive backtracking.
Since the computational costs of pencil marks are cheap, we do not includes pencil marks to difficulty.
We define three kinds of properties for Sudoku puzzles, depth, normal width, and average width, respectively.
The depth is the smallest number of recursive backtracking to solve a puzzle.
If a puzzle is solved only with pencil marks, then the depth of the puzzle is zero.
The width is the number of recursive backtracking required to confirm that the puzzle
has a unique solution.
It is possible that there are two or more empty cells which have the same number of pencil marks.
In Fig.~\ref{fig_pm}, the smallest number of pencil marks in cells is two,
and there are four cells which have two pencil marks.
A value of width depends on a choice of cells to perform recursive backtracking.
We define two kinds of width, the normal width and the average width.
\textbf{Normal width}:
If there are two or more candidates of cells, chose a cell in top-left order.
The choice of cells is deterministic, and the value of width is determined uniquely.
\textbf{Average width}:
If there are two or more candidates of cells, chose a cell randomly.
The value of the average width is stochastic.

The recursive backtracking process constructs a tree structure.
The root node of the tree is the given puzzle.
When a empty cell is chosen for recursive backtracking,
then the number of pencil marks in the cell corresponds to the number of edges of the nodes.
Each child node corresponds to a grid with an assumed number in the chosen cell.
The tree is constructed recursively for children.
Since we have a choice of cells to perform recursive backtracking, there are many possibility
for the tree structure. The depth is the shortest path from the root node to
the node describing answer between all possible trees. The width is a number of nodes in the tree graph.
Computational cost to search the solution is proportional to width of the graph.
Therefore, we adopt width as difficulty of Sudoku puzzles.
Since the value of the normal width depends on orientation of a puzzle,
the average width is more appropriate for difficulty of Sudoku puzzles
than the normal width.
Since a number of possible trees increases exponentially as a number of recursive backtracking increases,
it is impossible to enumerate all possible trees.
Therefore, we estimate the average width by MC sampling.
In the present manuscript, we calculate average width for each puzzle
from 100 independent samples.

\section{Method} \label{sec_method}

\subsection{Model}

Suppose a solution of a puzzle, \textit{i.e.}, all of cells are filled with numbers, is given.
To create a puzzle, we have to remove some of numbers from the grid.
The rest numbers are hints of this puzzle.
We label cells from $1$ to $81$, and  describe the state of $i$-th cell by a \textit{spin} $s_i$;
the $i$-th cell is empty when $s_i = 0$ and the cell keeps the number of the answer when $s_i = 1$.
A set of spin configuration $\{s_i\}$ denotes a puzzle.
We define an Ising spin-glass like Hamiltonian of this system as
\begin{equation}
H(\{s_i\}) = -J U(\{s_i\}) + h \sum_{i=1}^{81} s_i, \label{eq_hamiltonian}
\end{equation}
where $U$ denotes the internal energy given by the configuration $\{s_i\}$,
$J (>0)$ is the interaction energy, and $h (>0)$ is the amplitude of the external field, respectively.
The interaction energy $U$ is defined so that the energy of the system decreases
when the puzzle is more difficult.
We define two kinds of internal energy, depth energy $U_\mathrm{d}$ and 
width energy $U_\mathrm{w}$.
Consider a puzzle which depth is $d$ and normal width is $w$.
Then the depth and width energies are defined to be
\begin{eqnarray}
U_\mathrm{d} &=& d, \label{U_depth} \\
U_\mathrm{w} &=& \log(w). \label{U_width}
\end{eqnarray}
Since $J>0$, the energies decrease as depth or width increase.
We adopt logarithm for the width energy since width increases exponentially
as a number of recursive backtracking increases.
The energies depend on spin configuration $\{s_i\}$, but 
the relation between them is highly complicated.
While one can easily calculate the energies from the given spin configuration,
it is almost impossible to find the grand state, \textit{i.e.},
to find the spin configuration which describes the most difficult puzzle
for the given solution of the puzzle~\cite{NP}.
This property is similar to that of spin-glass models.

The second term in the right-hand side of Eq.~(\ref{eq_hamiltonian}) increase
energy when a number of hints of a puzzle increases. This term corresponds to an
external field.
While we have to minimize only the internal energy for the purpose to search a difficult puzzle,
we added this term since to apply a bias to reduce a number of hints.
Since the spins are all-up in the initial states, it is difficult to decrease energy without this term.
Our purpose is to create a puzzle width a large value of average width.
But we find that it is difficult to obtain large values of width from the initial state where all spins are up.
Therefore, we first perform MC simulations in the depth-first order,
then we switch to that in the width-first order.

In order to update a configuration of spins, we adopt the Markov Chain Monte Carlo method.
Choose a spin randomly and flip it with the Metropolis criterion with the Boltzmann weight~\cite{Metropolis1953}.
Note that, a spin-flip from down to up always increases energy, and that from up to down
always decreases energy.
If a new configuration has two or more solutions, then the trial is rejected.

\subsection{Adjusting Temperature Set}

Since the model is similar to Ising spin-glass models, there are many local minima
in the energy land scape. In order to search lower-energy states efficiently,
we adopt replica-exchange Monte Carlo (REMC) method~\cite{Hukushima1996,Swendsen1986},
which is also called Parallel Tempering method.
The REMC method is proposed by Hukushima and Nemoto in order to
study hardly-relaxing systems such as spin-glass systems efficiently.
In the REMC method, many replicas sharing the identical Hamiltonian
are simulated simultaneously and temperatures of replicas are sometimes exchanged.
The REMC method requires a tuned set of temperatures to work efficiently.
The set should includes a temperature which is high enough to
escape from any local minima and a temperature which is low enough
to search the grand state.
Additionally, a number of temperatures should be sufficient so that
exchange ratios between adjacent temperatures are high enough.

Usually, a temperature set is determined by preliminary simulations
and the set is fixed throughout simulations.
However, many temperatures, and consequently, many replicas
are required for Sudoku problems since the range of energy is wide.
Therefore, we adjust the temperature set simultaneously throughout simulations
to keep exchange ratios between replicas.
While we cannot obtain the canonical ensemble without a fixed set of temperatures,
it is not problem since we are interested only in the configuration having the lowest energy,

In order to obtain a temperature set which achieves same exchange ratio between neighboring
temperatures, the following procedure is proposed~\cite{Hukushima1996, Kerler1994}.
\begin{eqnarray}
\beta_1^{n+1} &=& \beta_1^{n}, \\
\beta_m^{n+1} &=& \beta_{m-1}^{n+1}  + 
 \frac{p_m^n}{c} \left(\beta_{m}^{n}  - \beta_{m-1}^{n} \right) \quad (2 \leq m), \\
c &\equiv & \frac{1}{M-1} \sum_{m=1}^{M} p_m^n,
\end{eqnarray}
where $M$ is a number of replicas, $\beta_m^{n}$ is the inverse temperature
of the $m$-th replica at the $n$-th exchange of temperature, and $p_m^n$ is the acceptance ratio of exchange
between the $m$-th and the $(m+1)$-th replicas, respectively.
The value of $p_m$ is estimated from MC sampling between exchange.
After convergence, the acceptance ratios will share the identical value 
as $p_1 = p_2 = \cdots = p_{M-1}$.
However, with a large number of iterations,
the temperature set can converge into the following trivial state
\begin{eqnarray}
\beta_1 &=& \beta_2 =  \cdots = \beta_M,\\
p_1 &=& p_2 =\cdots  = p_{M-1} = 1
\end{eqnarray}
In order to prevent the trivial convergence,
we determine a desired value of acceptance ratio $\bar{p}$ as follows,
\begin{eqnarray}
\beta_1^{n+1} &=& \beta_1^{n}, \\
\beta_m^{n+1} &=& \beta_{m-1}^{n+1}  + 
 \frac{p_m^n}{\bar{p}} \left(\beta_{m}^{n}  - \beta_{m-1}^{n} \right). 
\end{eqnarray}
Let $N_s$ is a number of MC steps between exchange processes,
\textit{i.e.}, a number of samples to estimate the acceptance ratios $\{p_m^n\}$.
If energy difference between adjacent replicas is extremely large,
then the acceptance ratio between the replicas becomes extremely small.
Then the acceptance ratio $p_m^n$ is estimated to be zero.
Once it happens, we have $\beta_m = \beta_{m+1}$ throughout the simulations
which means that the number of replicas virtually decreases.
In order to avoid the above, we adopts $1/N_s$ instead of $p_m^n$ if
the exchange is not performed between $m$-th and $(m+1)$-th replica in $N_s$ MC steps.
Finally, we obtain the following procedure to adjust temperatures as,
\begin{eqnarray}
\beta_1^{n+1} &=& \beta_1^{n}, \\
\beta_m^{n+1} &=& \beta_{m-1}^{n+1}  + 
c_m^n \left(\beta_{m}^{n}  - \beta_{m-1}^{n} \right),  \\
c_m^n &\equiv& \frac{\max\{p_m^n, 1/N_s\}}{\bar{p}}.
\end{eqnarray}
The above procedure guarantees that the all temperatures have different value
and a temperature increases when some of replicas are trapped in a local minimum.
We choose the initial set of temperatures as
\begin{equation}
\beta_m^{0} = \beta_1 + (\beta_M - \beta_1)\frac{m-1}{M-1}.
\end{equation}
The highest temperature $\beta_1$ is fixed throughout simulations.

\subsection{Details of Simulations}

From preliminary simulations, we adopt the interaction energy $J = 100$
and the external field $h = 1$ for both depth-first and width-first order calculations.
First we create a solution of Sudoku puzzle randomly, then
we perform the depth-first order search with the simple MC simulation with $\beta = 0.05$.
After we find a puzzle with depth larger than 8, we perform the width-first search with the REMC method.
We choose a number of MC steps between temperature exchange to be $N_s = 100$.
The highest temperatures is set to be $\beta_1 = 0.01$ which is high enough to escape any local minim.
The desired acceptance ratio for exchange $\bar{p}$
and a number of replicas $M$ are chosen to be $\bar{p} = 0.8$ and $M=10$, respectively.
The most time consuming part of this simulation is to calculate energy.
Therefore, we adopt the normal width for internal energy in Eq.~(\ref{U_width})
instead of the average width to save computational time.
We list up the candidates of hard puzzles from simulations, and determine the hardest one
by calculating average width for each candidate.
Since computational costs strongly depend on temperature,
it is inefficient to perform parallel computation for REMC due to load imbalance.
Therefore, we assign all replicas to one process, \textit{i.e.}, a simulation of each replica is performed serially.

\section{Results}
\label{sec_results}

\begin{figure}
\begin{center}
\includegraphics[width=9cm]{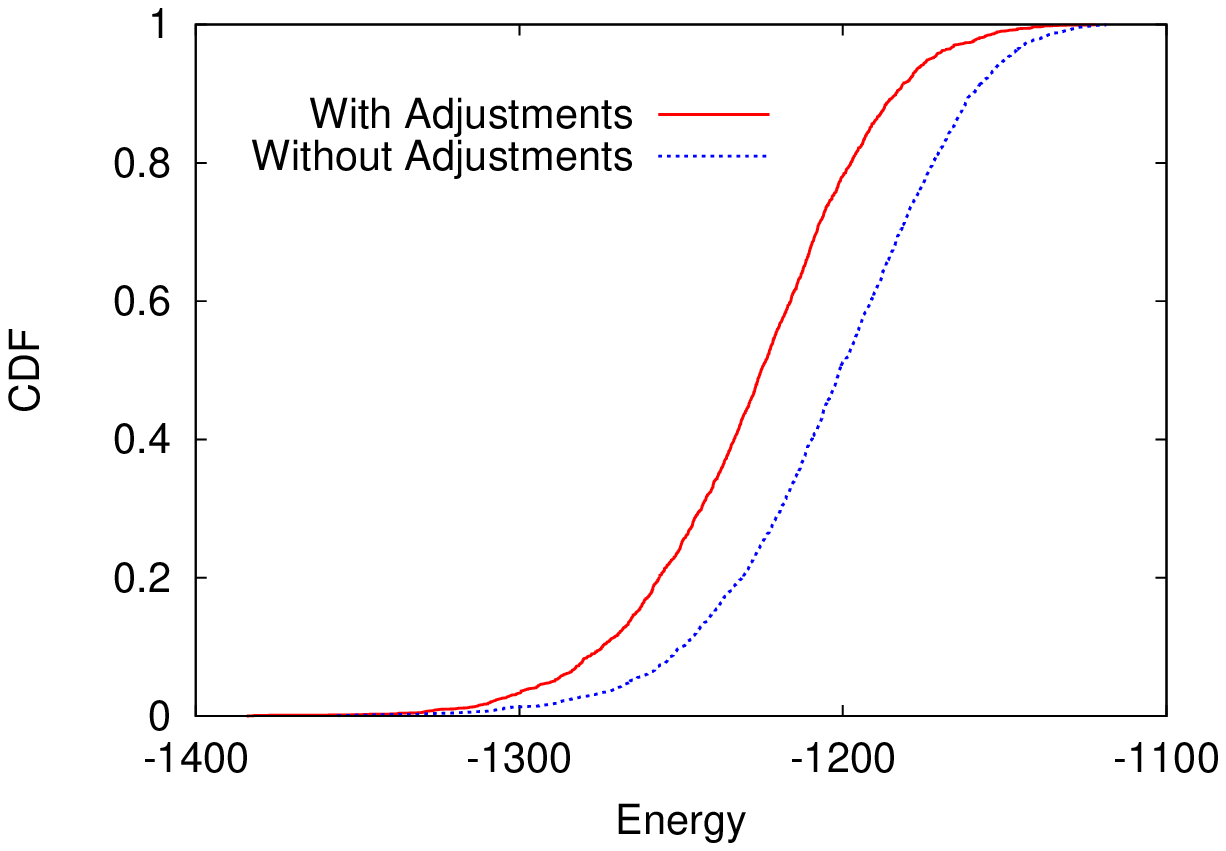}
\end{center}
\caption{(Color online)
Cumulative distribution functions of the minimum energy found by simulations
with and without adjustments of temperatures.
2048 independent runs are investigates for both cases.
The runs with adjustments of temperatures can find lower energies 
than those found by runs without adjustments of temperatures.
}\label{fig_hist}
\end{figure}

\begin{figure}
\begin{center}
\includegraphics[width=8cm]{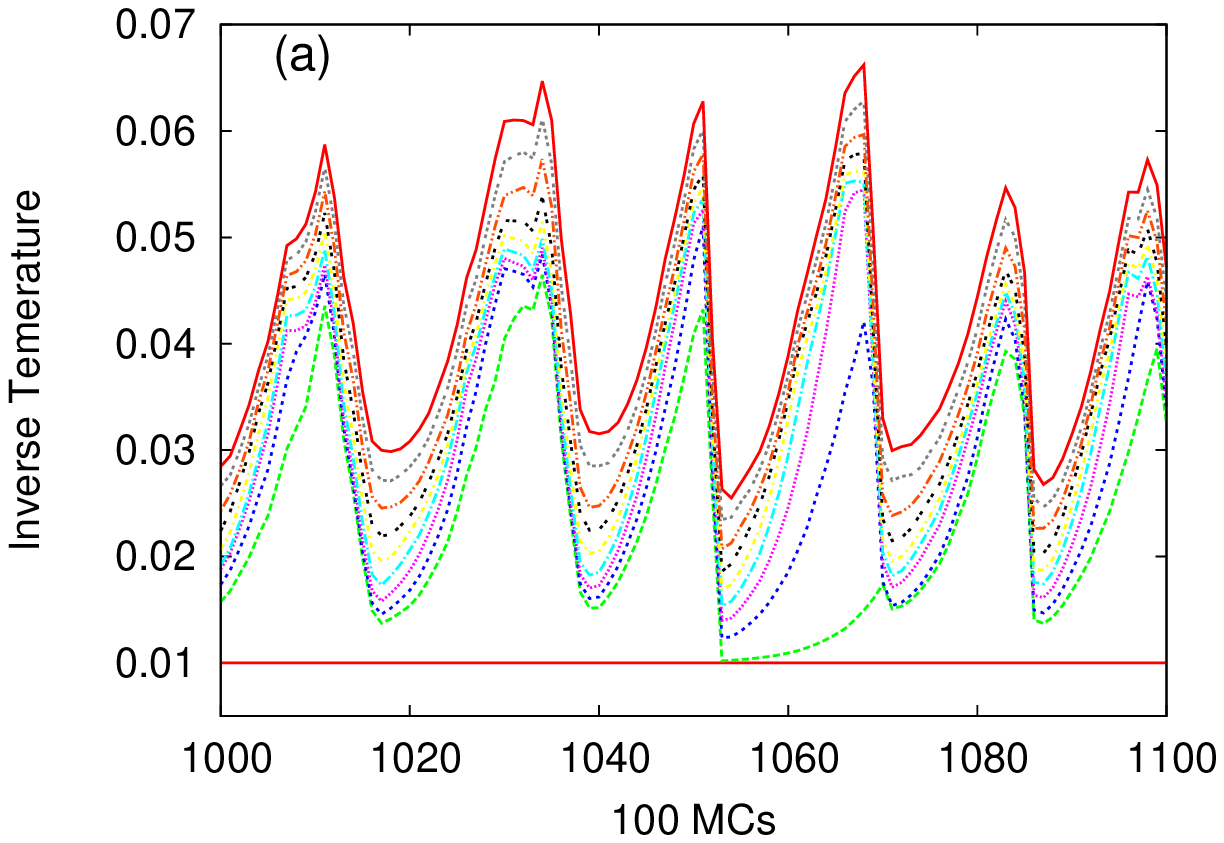}
\includegraphics[width=8cm]{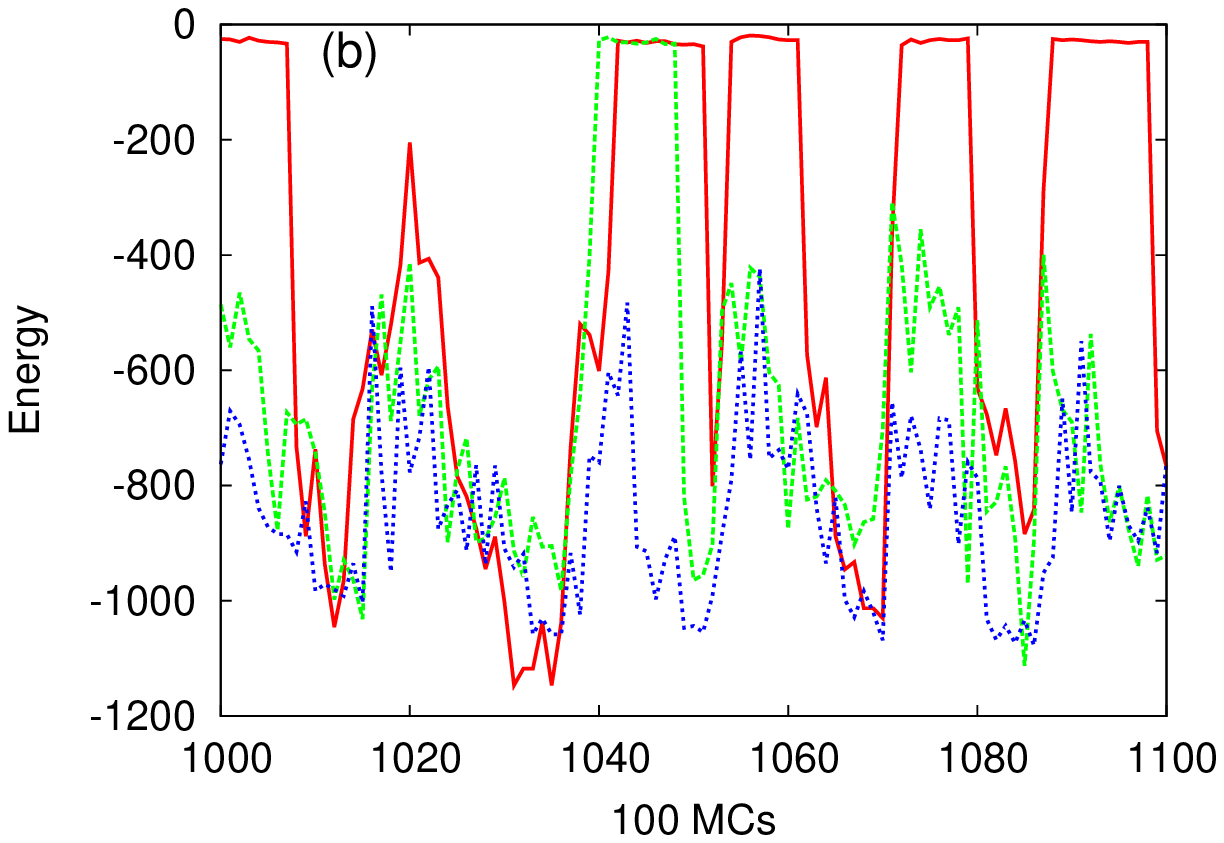}
\end{center}
\caption{(Color online) Typical time evolutions of a temperature set (a) and 
energies of replicas (b). While ten temperatures are shown, 
three energies out of ten are shown for the visibility.
The highest temperature, \textit{i.e.}, the lowest inverse temperature, is fixed to be $\beta_1 = 0.01$
throughout simulations.
One can see that the temperatures oscillate.
Thanks to the oscillation, the replicas can escape from local minima and visit both high- and low-energy states.
}\label{fig_results}
\end{figure}

In order to investigates the efficiency of the simultaneous temperature adjustments,
we perform simulations with and without the adjustments.
Each simulation contains 10 replicas, and 2048 independent samples are investigated
for simulations with and without adjustments.
We adopt the trivial parallelization, \textit{i.e.}, simulations are performed
independently with different seeds of random numbers.
Computations are carried on SGI Altix ICE 8400EX at the Institute for Solid State Physics, the University of Tokyo.
Computational time is 24 hours for each run and the total amount of
the computational time is about 11 CPU-core-years.

The cumulative distribution function (CDF) of the minimum energies found by the simulations
are shown in Fig.~\ref{fig_hist}.
The CDF $P(E)$ is a probability that the lowest energy found by each process is smaller than $E$.
It shows that the simulations with temperature adjustments is more efficient
than that without adjustments.

Typical time evolutions of temperatures and energies of a run with the temperature
adjustments are shown in Fig.~\ref{fig_results}.
The reason that temperatures oscillate is as follows. 
When one replica is trapped into a local minimum,
then acceptance ratio of exchange temperature between the replica and its neighbor decreases.
It increases temperature of the replica.
After the replica escapes from the local minimum, then the acceptance ratio
increases and the temperature decreases again.
From the variation rage of the lowest temperature, we estimate that
over 35 replicas are necessary to keep the desired acceptance ratio without the temperature adjustments,
while we used only 10 replicas. It means that the simultaneous temperature adjustments work efficiently.
Figure~\ref{fig_results}~(b) shows that
the energies of replicas fluctuate from high-energy states to low-energy states
which means that the exchange of temperatures works efficiently.

\begin{figure}
\begin{center}
\includegraphics[width=6cm]{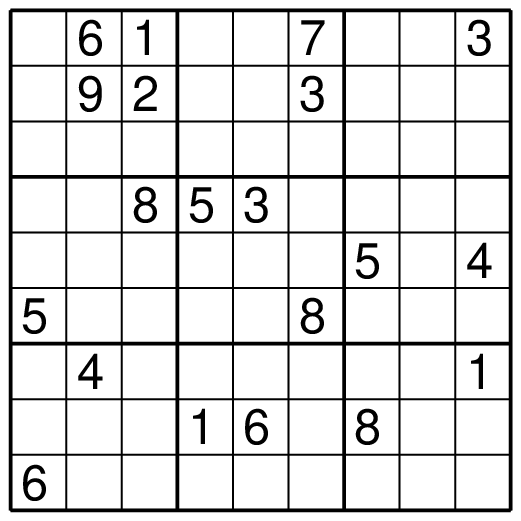}
\end{center}
\caption{
The hardest puzzle obtained by simulations.
Its depth is 10, normal width is 183530,
and average width is 100571 $\pm$ 1198.
}\label{fig_watanabe}
\end{figure}

We obtain 2048 candidates of difficult puzzles from the simulations with the simultaneous
temperature adjustments.
We calculate an average width for each candidates, and determine
the hardest one which is shown in Fig.~\ref{fig_watanabe}.
The hardest puzzle's depth is $10$ and average width is 100571 $\pm$ 1198, respectively.
This is about 44 times more difficult than Inkara2012.
If one adopts only pencil marks and recursive backtracking techniques to solve this puzzle,
then about 50000 times recursive backtracking is necessary to solve it, and therefore,
it is almost impossible to solve by hand.

\section{Summary and Discussion}
\label{sec_summary}

We define the Hamiltonian describing the difficulty of Sudoku puzzles.
Then creating difficult puzzles reduces to minimizing the energy defined by the Hamiltonian.
We perform the REMC method to minimize the Hamiltonian, and succeed to
create a Sudoku puzzle which is much harder than Inkara2012 in our definition of difficulty.
To our best knowledge, this is the world hardest puzzle ever created.
While the REMC is a method to obtain canonical ensemble of different temperatures simultaneously,
we propose the REMC with simultaneous adjustments of temperature which does not guarantee 
canonical ensembles.
The results presented in this manuscript demonstrates
that the REMC method is useful not only for physical problems,
but for general optimization problems.

A definition of difficulty strongly depends on solving algorithms.
We adopt only two techniques, pencil marks and recursive backtracking,
while there are many kinds of techniques to solve Sudoku puzzles.
Therefore, the difficulty defined in the present manuscript can be
different from the actual feeling.
But once the definition of the difficulty is given,
then difficult puzzles can be created in that definition
since the method proposed in the present manuscript is general and
it is independent of a definition of difficulty.

While we adopt the REMC method here, there are other 
optimization methods, such as genetic algorithm or simulated annealing, and so forth.
It is one of the further issues to compare such algorithms with REMC.
Since we are interested only in the lowest energy state, 
it is not necessary to achieve canonical ensemble for given temperature.
Therefore, we have a choice of the transition probability.
While we adopt the Boltzmann weight both for MC and REMC,
it is possible to accelerate finding lower energy states by adopting
general transition probability~\cite{Nishimori1998}.

In the present manuscript, we create a difficult Sudoku puzzle.
It is not always true that a difficult puzzle is interesting one.
However, if we can define a quantity describing how interest a puzzle is,
then we can create interesting puzzles by adopting the similar method proposed here.

The programs used in the present manuscript are published as  open source software~\cite{source}.

\section*{Acknowledgements}
The computations were carried out by using facilities of the
Supercomputer Center, Institute for Solid State Physics, University of Tokyo.
We would like to thank N. Kawashima, S. Todo, and T. Okubo for helpful discussions.


\begin{thebibliography}{99}
\bibitem{NP} Similar to solving Sudoku puzzles, it is expected that to determine the grand state
of Hamiltonian in Eq.~(\ref{eq_hamiltonian}) for the given solution would be also NP-compete.
However, it is not proven yet to our best knowledge.
\end{thebibliography}
\end{document}